\documentclass[letterpaper,twocolumn,10pt]{article}
\usepackage{usenix-2020-09}

\usepackage{tikz}
\usepackage{amsmath}

\usepackage{filecontents}
\usepackage{comment}
\usepackage{enumitem}

\usepackage{listings}
\usepackage{xcolor}
\usepackage[linesnumbered]{algorithm2e}

\definecolor{codegreen}{rgb}{0,0.6,0}
\definecolor{codegray}{rgb}{0.5,0.5,0.5}
\definecolor{codepurple}{rgb}{0.58,0,0.82}
\definecolor{backcolour}{rgb}{255,255,255}

\lstdefinestyle{mystyle}{
    backgroundcolor=\color{backcolour},   
    commentstyle=\color{codegreen},
    keywordstyle=\bfseries\color{black},
    numberstyle=\tiny\color{codegray},
    stringstyle=\color{black},
    basicstyle=\ttfamily\small,
    breakatwhitespace=false,         
    breaklines=true,                 
    captionpos=b,                    
    keepspaces=true,                 
    numbers=left,                    
    numbersep=2pt,                  
    showspaces=false,                
    showstringspaces=false,
    showtabs=false,                  
    tabsize=2,
    escapeinside={\%*}{*)}
}
\newcommand{\tool}[0]{\mbox{\textsc{BinGo}}}

\begin{document}

\date{}

\title{\Large \bf \tool{}: Pinpointing Concurrency Bugs in Go via Binary Analysis}


\author{
{\rm Chongxin Zhong *}\\
North Carolina State University
\and
{\rm Qidong Zhao *}\\
North Carolina State University
\thanks{Chongxin and Qidong contributed equally to this work and should be considered co-first authors.}
\and
{\rm Xu Liu}\\
North Carolina State University
} 

\maketitle

\begin{abstract}
Golang (also known as Go for short) has become popular in building concurrency programs in distributed systems. As the unique features, Go employs lightweight Goroutines to support highly parallelism in user space. Moreover, Go leverages channels to enable explicit communication among threads. However, recent studies show that concurrency bugs are not uncommon in Go applications. Pinpointing these concurrency bugs in real Go applications is both important and challenging. Existing approaches are mostly based on compiler-aided static or dynamic analysis, which have two limitations. First, existing approaches require the availability and recompilation of the source code, which work well on testing rather than production environments with no source code available for both applications and external libraries. Second, existing approaches work on pure Go code bases only, not programs mixed with Go and other languages. To address these limitations, we develop \tool{}, the first tool to identify concurrency bugs in Go applications via dynamic binary analysis. \tool{} correlates binary execution with Go semantics and employs novel bug detection algorithms. 
\tool{} is an end-to-end tool that is ready for deployment in the production environment with no modification on source code, compilers, and runtimes in the Go eco-system. Our experiments show that \tool{} has a high coverage of concurrency bugs with no false positives. We are able to use \tool{} to identify concurrency bugs in real applications with moderate overhead.
\end{abstract}

\section{Introduction} \label{introduction}

Golang (short as Go) is a modern open-source programming language created by Google in 2007. 
Go has become increasing popular recently as it supports building fast, reliable, and efficient software at scale~\cite{go-www}. A variety of microservices running on cloud adopt Go as the default language for transaction processing with high concurrency.
Popular software written in Go includes CockroachDB~\cite{cockroach-www}, Dropbox~\cite{dropbox-www}, AresDB~\cite{aresdb-www}, gRPC~\cite{grpcgo-www}, Docker~\cite{docker-www}, Kubernetes~\cite{kubernetes-www}, and many more. Leading companies, such as Google, Facebook (Meta), Uber, Microsoft and many others form an active community to support Go.

As the unique features,
Go facilitates programming concurrency by introducing {\tt go} keyword, which can be placed as a function prefix to indicate this function can be run asynchronously. Functions with {\tt go}
as prefix are called Goroutines. Application developers can enjoy the concurrency by creating multiple Goroutines. Compared to traditional OS threads (e.g., pthread), Goroutines are lightweight because the Go runtime system context switches Goroutines in the user space, with no kernel involved. There are two mechanisms to communicate data between Goroutines: (1) shared memory, which is similar to traditional threads, and (2) channels, which support explicit message passing.

Like other parallel programs, concurrency bugs are not uncommon in Go applications. Concurrency bugs occur due to the misuse of synchronization across concurrent Goroutines. Concurrency bugs can show the symptoms of both blocking and non-blocking execution. The blocking bugs, similar to deadlocks, result in no progress of Goroutines, while non-blocking bugs are similar to data races, resulting in incorrect results. We focus our study on blocking bugs in this paper. Recent studies~\cite{10.1145/3297858.3304069, 9370317} show that blocking bugs in Go applications are not related to only mutex primitives but also channels. These blocking bugs involve multiple synchronization contexts buried deep in code bases, which are difficult to identify.

A variety of tools have been developed to identify concurrency bugs in Go, which mostly use two techniques---static and dynamic analysis. The static analysis leverages compilers to analyze Go source code without the need of executing the code. However, static analysis has a limited coverage in identifying concurrency bugs with a small set of predefined patterns. Moreover, static analysis can introduce many false positives or negatives due to the imprecise analysis raised by aliases or pointers. Such imprecision can be quickly accumulated when analysis scope enlarges in real Go applications. 

To overcome the limitations of static analysis, dynamic analysis~\cite{goruntime,goleak} monitors Go program execution with additional runtime information for accurate analysis on aliases and pointers. With representative or fuzzed inputs, dynamic analysis minimizes false positives and negatives with high bug coverage. However, existing dynamic approaches require the modification of Go compilers or runtimes to collect necessary data. Such requirement largely impedes these tools from the adoption in the production environment with the off-the-shelf Go software stack.

In this paper, we propose \tool{}, a new bug detector for Go applications to address challenges in existing static and dynamic tools. The key idea of \tool{} is to decouple the data collection and analysis: \tool{} collects the runtime information with binary analysis and analyzes concurrency bugs with integrating Go semantics.
The binary analysis allows \tool{} to work for unmodified, off-the-shelf Go ecosystem and semantic-aware analysis enables \tool{} to accurately pinpoint concurrency bugs with no false positives and provide intuitive guidance for big fixes. 

\tool{} advances existing approaches in three aspects. First,  \tool{} can handle legacy or commercial Go code packages that are only available as binary code. Second, \tool{} can analyze not only pure Go applications but also code hybrid with Go and other languages, such as C via {\tt cgo}. Third, \tool{} requires no code change, specific compilation options, or code recompilation, so \tool{} is more suitable for practical deployment. 

In the rest of this section, we show a motivating example, highlight the challenges \tool{} addresses, and summarize our contributions.

\begin{figure}[tp]
\centering
\begin{lstlisting}[language=Go]
go func() { // Goroutine1
	%*\color{red}defer close(wbs.donec)*)
	for wb := range wbs.updatec {
		wbs.coalesce(wb) // Lock, Unlock
	}
}()
func (wbs *watchBroadcasts) coalesce(wb *watchBroadcast) {
	%*\color{red}wbs.mu.Lock()*)
	%*\color{red}wbs.mu.Unlock()*)
}
func (wbs *watchBroadcasts) stop() { // Goroutine2
	%*\color{red}wbs.mu.Lock()*)
	%*\color{red}defer wbs.mu.Unlock()*)
	close(wbs.updatec)
	%*\color{red}<-wbs.donec*)
}
\end{lstlisting}
\caption{A channel-mutex blocking bug in a real Go application etcd~\cite{etcd-www}. We omit codes that are not related to synchronization for simplicity.}
\label{code1}
\end{figure}

\paragraph{Motivation}

Figure~\ref{code1} shows an example Go program from etcd~\cite{etcd-www} with concurrency bugs with both mutex locks and channels involved. With manual code analysis, we find that with one execution order, concurrency bugs happen. Goroutine2 first acquires the mutex lock {\tt wbs.mu} at line 12 and gets blocked on receiving messages in channel {\tt wbs.donec} at line 15. At the meanwhile, Goroutine1 calls function {\tt wbs.coalesce()}, where it also tries to acquire the mutex lock {\tt wbs.mu} at line 8 before closing {\tt wbs.donec} at line 2. Thus, Goroutine1 and Goroutine2 are both blocked with no progress. Manual analysis of such concurrency bugs, which require understanding the program semantics of synchronization primitives and exploring every execution order of these primitives, is both tedious and error-prone.

\paragraph{Challenges in analyzing Go binaries}
        The challenges of enabling binary analysis for Go applications reside in the correlation between the low-level binary instruction behaviors and the high-level Go semantics, such as operations on mutex locks and channels as shown in Figure~\ref{code1}. 
        Analyzing the native binary traces cannot either accurately identify concurrency bugs or provide intuitive guidance for bug fixes.  To understand whether concurrency bugs exist, \tool{} needs to interpret synchronization primitives (mutexes and channels) and understand the behaviors of Goroutines (e.g., creation, scheduling, and destruction). To provide intuitive guidance, \tool{} needs to understand the logical call paths rather than the native call path constructed from the binary. \tool{} is able to address these challenges with pure binary analysis by leveraging the open-source Go compilers and runtime systems. Details are in Sections~\ref{methodology} and~\ref{implementation}.

\paragraph{Contributions}
\tool{} advances the state of the arts with the following contributions.
\begin{itemize}[leftmargin=*]
    \item \tool{} is the first tool that analyzes binary of Go applications. \tool{} supports a hierarchy of trace points via investigating Go runtime binaries to associate binary execution with Go semantics.
    
    \item \tool{} employs a novel predictive algorithm with the combination of online and offline analysis to detect various concurrency bugs in Go applications that involve Goroutines, channels, and synchronization primitives.
    
    \item \tool{} adopts many optimization techniques to reduce its overhead, enhance its scalability, and integrating the bug report in use-friend GUI for applicability. \tool{} is an end-to-end bug detector running in off-the-shelf Go ecosystem. \tool{} will be open source upon the paper acceptance.
    \item \tool{} has been evaluated extensively with well-known Go benchmarks. \tool{} shows high bug coverage, affordable overhead, and deep insights to guide bug fix. Furthermore, \tool{} is deployed to analyze real Go applications; \tool{} can identify concurrency bugs and provide intuitive guidance for the fix.
\end{itemize}

\paragraph{Paper organization}
The remaining paper is organized as follows. Section~\ref{related_work} reviews the existing approaches and distinguishes \tool{}. Section~\ref{methodology} describes the methodology of \tool{}. Section~\ref{implementation} elaborates on the implementation details of \tool{}. Section~\ref{evaluation} evaluates the coverage and overhead of \tool{}. Section~\ref{case_stuties} studies several cases of concurrency bugs in real Go applications. Section~\ref{discussion} shows some discussions. Finally, Section~\ref{conclusions} presents some conclusions of the paper and previews the future work.

\section{Related Work} \label{related_work}

Concurrency bug detection has been extensively studied recently~\cite{10.5555/2387880.2387902,10.1145/1950365.1950395,10.1145/3341301.3359638,10.5555/3277355.3277436,10.1145/3037697.3037735}. However, these approaches cannot directly apply to Go applications due to the lack of support for Go language features. 
In this section, we only review the most related approaches to \tool{}, including concurrency bug detection in Go and binary analysis for software bugs.

\subsection{Concurrency Bug Detection in Go}

Staticcheck~\cite{staticcheck} and Vet~\cite{vet} analyze Go source code statically to identify concurrency bugs. These two approaches, based on pattern recognition, have limited coverage of bug detection. Tools~\cite{gabet_et_al:LIPIcs:2020:13161, 10.1145/3093333.3009847, 8453195, 10.1145/2892208.2892232, 10.1145/3314221.3322484} based on model checking address the coverage limitation but incur tremendous overhead by investigating each input and modeling the execution of all synchronization primitives, which impedes them from applying to real Go applications.

GCatch~\cite{10.1145/3445814.3446756} is the state-of-the-art static bug detector for Go applications. It uses the Go compiler to analyze source code and build a constraint solver to detect blocking bugs. However, it (1) only detects channel-related bugs, (2) has limited analysis scope, and (3) yields imprecise analysis with many false positives. Section~\ref{evaluation} gives more quantitative comparison between \tool{} and GCatch. 

Moreover, due to the static analysis, these approaches
do not provide sufficient information, e.g., calling contexts, to help understand the root causes of concurrency bugs. Confirming and fixing these bugs require substantial manual efforts. Thus, dynamic analysis is proposed as an alternative solution, which \tool{} is also built on.

Go runtime by default can detect concurrency bugs in its Goroutine scheduler dynamically~\cite{goruntime}. When {\em all} Goroutines are blocked, the runtime reports a concurrency bug. However, this scheme only detects a small set of concurrency bugs; it cannot identify blocking that occurs among a subset of Goroutines. Goleak~\cite{goleak}, a dynamic tool developed by Uber, can detect Goroutines leaks. Similar to the Go runtime, Goleak can only detects a small set of concurrency bugs. 
It is worth noting that most concurrency bugs studied in this paper cannot be identified by these tools. Moreover, these tools, leveraging the Go compiler for code instrumentation, target applications written in Go only, but cannot handle applications with hybrid languages such as Go and C (programmed with {\em cgo}).  

Unlike existing approaches, \tool{} adopts dynamic binary analysis with no compiler involved. Moreover, \tool{} correlate the binary analysis with Go semantics to yield accurate bug detection with rich insights.

\subsection{Binary Analysis for Software Bugs}

The dynamic binary analysis has already become the indispensable component in the system software stack nowadays. Binary engines, such as Pin~\cite{Luk-Cohn-etal:2005:PLDI-pin}, DynamoRIO~\cite{dynamorio-www}, Dyninst~\cite{dyninst-www}, and Valgrind~\cite{valgrind-www}, can instrument any instruction of interest for performance characterization and bug detection.
The performance characterization tools include redundancy detection~\cite{redspy,loadspy,RVN,Chabbi:2012:DTP:2259016.2259033}, data locality measurement~\cite{ibs-ispass, Xiang:2013:HHO:2451116.2451153}, cache simulation~\cite{6114398}. The bug detection includes aforementioned concurrency bug analysis, taint analysis~\cite{Brumley:2006:TAG:1130235.1130359}, reverse engineering~\cite{Lin08automaticprotocol}, and execution replay~\cite{Narayanasamy:2005:BCR:1069807.1069994}.

However, these tools target applications written in native languages only. As discussed in Section~\ref{implementation}, these tools do not apply to Go programs, because they capture the raw measurement only, lacking of Go semantics. 

\section{Methodology} \label{methodology}

We describe the methodology of \tool{} to answer two research questions: (1) how to associate binary with Go semantics and (2) with the Go semantics, how to identify concurrency bugs in Go with high accuracy and coverage?

\subsection{Recovering Go Semantics in Binary} \label{recovering}

Since most concurrency bugs are related Go semantics, such as the operations on Goroutines and channels~\cite{10.1145/3297858.3304069}, it is important for \tool{} to identify these semantics when monitoring Go binary execution. However, Go binaries, like other binary languages, typically contain no program semantics, except symbol tables and debugging information if not stripped, which raises unique challenges for \tool{}.
Fortunately, The mainstream Go ecosystem developed by Google, including compiler and runtime, is open source. \tool{} is able to leverage this opportunity to recover Go semantics from binary analysis only. The key idea behind \tool{} is to recognize certain semantics via interpreting Go runtime source code. \tool{} then intercepts the binary by adding a hierarchy of tracepoints to uncover the semantics.

\begin{table}[t]
\scriptsize
\centering
\renewcommand\arraystretch{1}
\resizebox{.38\textwidth}{!}{%
\begin{tabular}{ll}
\hline
Semantic Name                  & Description              \\ \hline
$\emph{G}_{Create}$     & Goroutine creation       \\
$\emph{G}_{exit}$       & Goroutine exit           \\
$\emph{Main}_{start}$   & main.main function start \\
$\emph{Obj}_{create}$   & Object creation          \\
$\emph{M}_{l}$          & Mutex Lock               \\
$\emph{M}_{u}$          & Mutex Unlock             \\
$\emph{RWM}_{rl}$       & RWMutex RLock            \\
$\emph{RWM}_{ru}$       & RWMutex RUnlock          \\
$\emph{RWM}_{l}$        & RWMutex Lock             \\
$\emph{RWM}_{u}$        & RWMutex Unlock           \\
$\emph{Chan}_{create}$  & Channel creation         \\
$\emph{Chan}_{send}$    & Channel send             \\
$\emph{Chan}_{recv}$ & Channel receive          \\
$\emph{Chan}_{close}$   & Close Channel            \\
$\emph{Select}$         & Select statement         \\
$\emph{WG}_{add}$       & WaitGroup Add            \\
$\emph{WG}_{wait}$      & WatiGroup Wait           \\
$\emph{Ctx}_{create}$  & Context Creation         \\
$\emph{Ctx}_{cancel}$  & Context Cancel           \\
$\emph{Cond}_{wait}$    & Cond Wait                \\
$\emph{Cond}_{sig}$     & Cond Signal              \\
$\emph{Cond}_{bc}$      & Cond Broadcast           \\ \hline
\end{tabular} %
}
\caption{The necessary semantics in Go runtime for detecting concurrency bugs. The list can be expanded if more semantics are needed.}
\label{tab:actions}
\end{table}

\paragraph{Recognizing semantics in Go runtime}
Table~\ref{tab:actions} lists the main semantics \tool{} needs to recognize for identifying concurrency bugs. The semantics are represented as actions (e.g., Goroutine creation/destruction, lock acquisition/release, channel selection). Moreover, the objects handled by each action should be captured and assigned an ID to understand the correlation among various actions. Such objects include Goroutines, mutexes, and channels, etc. \tool{} outputs the function signatures related to these semantics.
It is worth noting that \tool{} needs to identify the Go version used in the system first and analyze its runtime source code once; all the following-up analyses in this system can reuse this output.


\begin{figure*}[t]
\centering
\includegraphics[width=1\textwidth]{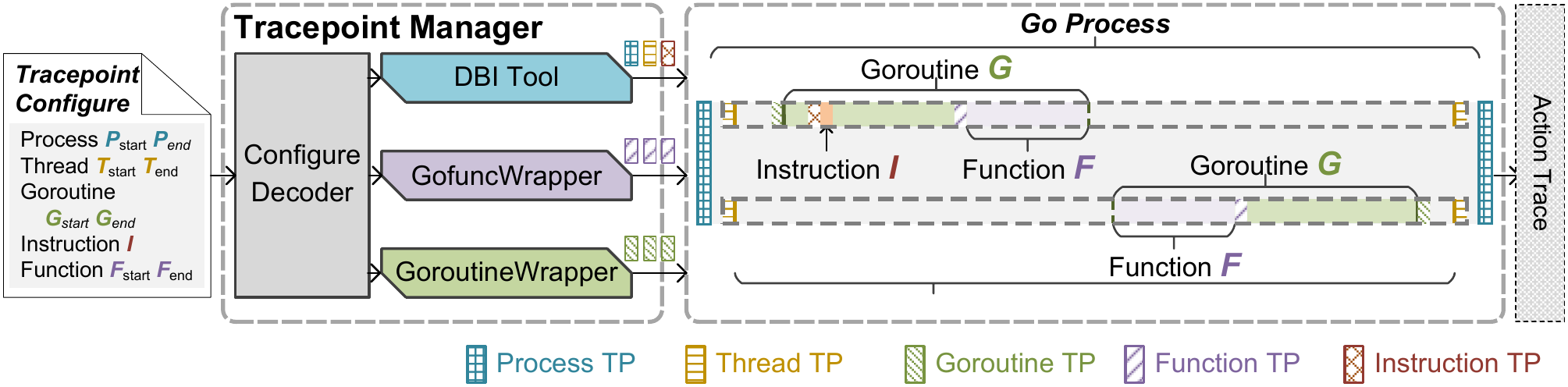}
\centering
\caption{Overview of tracepoint mechanism. The location of tracepoints are obtained via static analysis on Go runtimes. With the help of dynamic binary instrumentation (DBI) tool, tracepoints can be installed at different levels, such as instructions, functions, and Goroutines.}
\label{fig:tp-arch1}
\end{figure*}


\paragraph{Recovering semantics via tracepoints}
The semantics of an application can be composed of actions (e.g., Goroutine creation) and objects (e.g., mutex locks or channels with their IDs). \tool{} develops a mechanism that can freely arm tracepoints in Go binaries to gain actions and objects from the Go runtime.
Figure~\ref{fig:tp-arch1} overviews the tracepoint mechanism. The tracepoints can occur on the Goroutine level, function level, or the instruction level. At each tracepoint, \tool{} can collect calling context, memory information, and registers states. \tool{} then can activate a callback function (also known as tracepoint activation) that handles the information collected at the tracepoint and returns actions and objects. With these actions and objects, \tool{} can recover semantic details about the analyzed application for the blocking bug detection. 

\RestyleAlgo{ruled}
\begin{algorithm}[hbt!]
\caption{A high-level algorithm for detecting blocking bugs in \tool{}.}\label{alg1}
Online analysis for each Goroutine: \\
{
    \If{tracepoint is activated} {
        record the timestamp\\
        record the action\\
        record the lock or channel object
        and maintain them in a trace
    } 
    \If{blocking occurs} {
        report the problematic tracepoints
    }
}

Offline analysis on tracepoint trace of each Goroutine: \\
{
\If{blocking potentially occurs} {
    report the problematic tracepoints with prediction
}
}

\end{algorithm}

\subsection{Detecting Blocking Bugs}

The blocking bugs can happen among locks, channels, and other synchronization primitives. Prior work~\cite{10.1145/3297858.3304069} categorizes different blocking bugs in Go; \tool{} is able to detect all of them with a hybrid online and offline analysis, with the support of the tracepoints. 

Algorithm~\ref{alg1} shows a generic algorithm that \tool{} uses to handle each action and object. 
\tool{} begins the monitoring at when the tracepoint $Main_{start}$ is triggered, which is the entrance function of Go applications. 
In the remaining section, we show how \tool{} identifies different kinds of blocking bugs.

\subsubsection{Blocked Channel Operations}
When the tracepoints $Chan_{send}$, $Chan_{recv}$, or $Select$ are triggered, \tool{} records the context of these tracepoints. If the function enclosing these tracepoints does not return until the whole program execution finishes, \tool{} reports a blocked channel operation and uses the context information of these operations to associate the bug with the source code.

\subsubsection{Blocked WaitGroup}
\tool{} installs the tracepoints $WG_{add}$ and $WG_{wait}$ to track the number of the Goroutines in a WaitGroup,
If the number is not 0 at the end of program execution, meaning some Goroutines are blocked in WaitGroup. \tool{} reports the context of all $WG_{wait}$ tracepoints, which are the potential cause of the blocking.


\subsubsection{Blocked Cond}
For each {\tt Cond} variable, \tool{} keeps track of its operations. For a {\tt Cond} variable, if $Cond_{wait}$ is triggered in a Goroutine but no $Cond_{sig}$ or $Cond_{bc}$ triggered in other Goroutines before the program ends, \tool{} reports the $Cond_{wait}$ operation is blocked.

\subsubsection{Uncanceled Context}
Go programs use {\tt context} to pass values, cancellation signals, and deadlines across different Goroutines. Upon a {\tt context} creation tracepoint $Ctx_{create}$, \tool{} obtains the {\tt context} address and its {\tt cancel} function. 
This cancel function enables Goroutines blocked on this {\tt context} to receive messages from the function {\tt Done} for resuming execution. Thus, if a context is not canceled (i.e., $Ctx_{cancel}$ tracepoint not triggered) when the program finishes, \tool{} treats all the Goroutines waiting for this context as blocking bugs.

\subsubsection{Missing Unlock and Double Lock}
To reduce the runtime overhead and apply the prediction mechanism, \tool{} analyzes the tracepoints related to {\tt Mutex} offline to detect missing unlock and double lock in each Goroutine. Section~\ref{prediction} shows more algorithm details.

\subsubsection{Goroutine Leak}
\tool{} keeps track of the status of each Goroutine via tracepoints $G_{create}$ and $G_{exit}$. Upon $G_{create}$, the newly created Goroutine (except for the main Goroutine) is added to a list. At $G_{exit}$, the Goroutine is removed from that list. If the list is not empty after the entire program execution finishes, the Goroutines in the list may be leaked. However, it is possible that the program ends shortly after a Goroutine exits, $G_{exit}$ may not be triggered. This is not a Goroutine leak. Thus, \tool{} checks Goroutine leak with other blocked bugs: if a Goroutine has a blocked operation and does not exit before the program ends, \tool{} reports a leak.

\subsection{Predicting Blocking Bugs}
\label{prediction}
To enlarge the bug coverage, \tool{} identifies the blocking bugs due to deadlocks across mutex locks and channels by analyzing the trace, even with the blocking not occurring. We elaborate the algorithms \tool{} uses to predict the deadlocks among locks and channels.

\subsubsection{Mutex Deadlock}
\tool{} employs a dynamic deadlock detection mechanism, which is a variance of prior work~\cite{6718069}. Besides adapting the algorithm to Go semantics, \tool{} also extends the algorithm to support both traditional {\tt Mutex} locks and {\tt RWMutex} reader-writer locks in Go. 
For simplicity, we refer to both {\tt Mutex} and {\tt RWMutex} as mutex in the following texts. We define a lock tuple $\tau = \langle g, m, o, L \rangle$, which means that the Goroutine $g$ holds all the mutexes in the mutex set $L$ and tries to acquire the mutex $m$; $o$ represents the lock type: $M_l$, ${RWM}_{rl}$, or ${RWM}_{l}$.
We further define the lock sequence ($d$): a sequence of $n (n > 1)$ lock tuples, $d = \langle \tau_1, \tau_2... \tau_n \rangle$ where $\tau_i = \langle g_i, m_i, o_i, L_i \rangle$, $m_i \in L_{i+1}$, ..., $m_{k-1} \in L_k$, $g_i \neq g_j$, $L_i \cap L_j = \emptyset$ for $1 \leq i$, $j \leq n(i \neq j)$. Let the type of the lock operation of $m_i (1 \leq i < n)$ in $L_{i+1}$ be $p_{i+1}$ and the type of the lock operation of $m_n$ in $L_1$ be $p_1$. If $m_n \in L_1$, $\neg(o_i =$ ${RWM}_{rl}$ $\wedge$ $p_{i+1} = {RWM}_{rl})$, and $\neg(o_n =$ ${RWM}_{rl}$ $\wedge$ $p_1 =$ ${RWM}_{rl})$ for $1 \leq i < n$, a potential deadlock in the lock sequence $d$. Similarly, \tool{} also detects missing unlock and double lock in a lock sequence. The correctness of this algorithm is out of the scope. One can refer to the prior work~\cite{6718069} for more details.

\RestyleAlgo{ruled}
\begin{algorithm}[hbt!]
\caption{Channel-mutex deadlock prediction algorithm used by \tool{}.}\label{alg2}
\For{each $Chan_{op}$ = $Chan_{send}$ $\vee$ $Chan_{recv}$ $\vee$ $Chan_{close}$} {
    construct locked and unlocked mutex sets $L$ and $U$ in the same Goroutine
}
\For{each pair of $Chan_{opi}$, $Chan_{opj}$ $\in$ $Chan_{op}$ of the same channel} {
    \If{$Chan_{opi}$, $Chan_{opj}$ in different Goroutines} {
        \If{$Chan_{opi}$ = $Chan_{send}$ $\wedge$ $Chan_{opj}$ = $Chan_{recv}$} {
            \If{$L_i \cap L_j \neq \emptyset \vee L_i \cap U_j \neq \emptyset \vee U_i \cap L_j \neq \emptyset$} {
                report a deadlock
            }
        }
        \If{$Chan_{opi}$ = $Chan_{recv}$ $\wedge$ $Chan_{opj}$ = $Chan_{close}$} {
            \If{$L_i \cap L_j \neq \emptyset \vee L_i \cap U_j \neq \emptyset$} {
                report a deadlock
            }
        }
    }
}
\end{algorithm}

\subsubsection{Channel-Mutex Deadlock}
To predict whether blocking bugs exist due to channels, \tool{} collects the sequence of both mutex locks and  channels along with the timestamps. \tool{} then constructs the {\tt lock} and {\tt unlock} dependency of each channel operation and detects deadlocks if cyclic dependency exists. Algorithm~\ref{alg2} describes the algorithm elaborates on this algorithm.

\section{\tool{} Implementation} \label{implementation}
\begin{figure*}[t]
\centering
\includegraphics[width=1\textwidth]{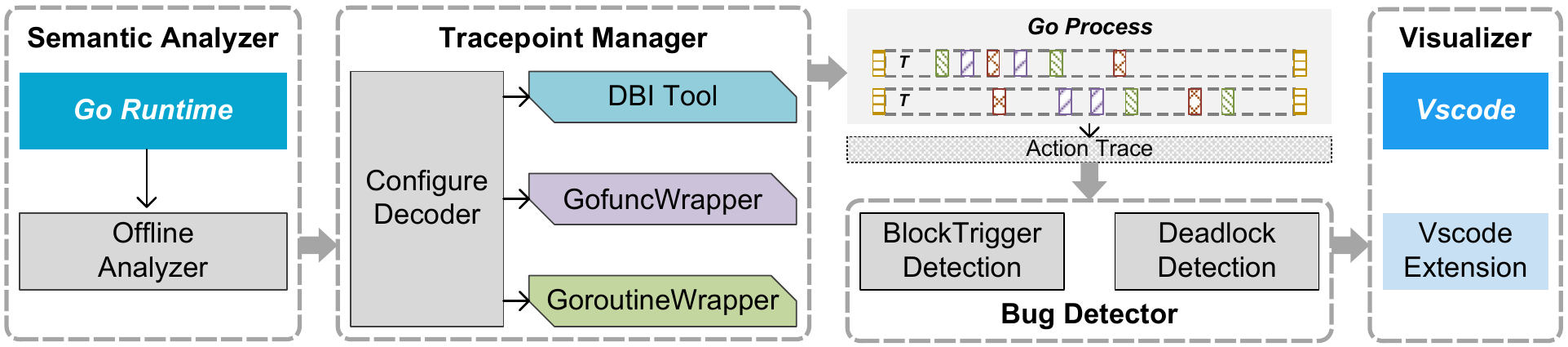}
\centering
\caption{Overview of the infrastructure of \tool{}. There are four components: semantic analyzer, tracepoint manager, bug detector, and visualizer. The tracepoint manage and bug detector use dynamic analysis; other components use static offline analysis.}
\label{fig:overview}
\end{figure*}

In this section, we elaborate on the implementation details of \tool{}. \tool{} is implemented in 6.4k lines of C++ code and maintained as an open-source project. 
Figure~\ref{fig:overview} overviews the structure of \tool{}. It mainly consists of $4$ components: (1) a semantic analyzer that extracts Go semantics, (2) a Tracepoint manager that activates necessary tracepoints according to Go semantics, (3) a bug detector that identifies blocking bugs, and (4) a visualizer that presents the analysis results in a user-friendly GUI.


\subsection{Semantic Analyzer}

\begin{table}[t]
\scriptsize
\centering
\renewcommand\arraystretch{1}
\resizebox{.38\textwidth}{!}{%
\begin{tabular}{ll}
\hline
Semantic Name                  & Function              \\ \hline
$\emph{G}_{Create}$     & runtime.newproc1       \\
$\emph{G}_{exit}$       & runtime.goexit0           \\
$\emph{Main}_{start}$   & main.main \\
$\emph{Obj}_{create}$   & runtime.newobject          \\
$\emph{M}_{l}$          & sync.(*Mutex).Lock               \\
$\emph{M}_{u}$          & sync.(*Mutex).Unlock             \\
$\emph{RWM}_{rl}$       & sync.(*RWMutex).RLock            \\
$\emph{RWM}_{ru}$       & sync.(*RWMutex).RUnlock          \\
$\emph{RWM}_{l}$        & sync.(*RWMutex).Lock             \\
$\emph{RWM}_{u}$        & sync.(*RWMutex).Unlock           \\
$\emph{Chan}_{create}$  & runtime.makechan         \\
$\emph{Chan}_{send}$    & runtime.chansend             \\
$\emph{Chan}_{recv}$ & runtime.chanrecv          \\
$\emph{Chan}_{close}$   & runtime.closechan            \\
$\emph{Select}$         & runtime.selectgo         \\
$\emph{WG}_{add}$       & sync.(*WaitGroup).Add            \\
$\emph{WG}_{wait}$      & sync.(*WaitGroup).Wait           \\
$\emph{Ctx}_{create}$  & context.(*cancelCtx).cancel         \\
$\emph{Ctx}_{cancel}$  & context.(*cancelCtx).cancel           \\
$\emph{Cond}_{wait}$    & sync.(*Cond).Wait                \\
$\emph{Cond}_{sig}$     & sync.(*Cond).Signal              \\
$\emph{Cond}_{bc}$      & sync.(*Cond).Broadcast           \\ \hline
\end{tabular} %
}
\caption{The semantics and their related functions in Go runtime. The list can be expanded if more semantics are needed.}
\label{tab:functions}
\end{table}

This component identifies all the interesting semantics in Go runtime. To guarantee the high accuracy, we manually analyze the Go runtime source code to recognize all the functions that related to the semantics. It is worth noting that, the manual source code analysis only needs to be done once for each Go release version. 
Table~\ref{tab:functions} shows the semantics and their related functions in Go v1.17.3. This table (also known as semantic table) is maintained in a text file for later usage.

\subsection{Tracepoint Manager}

The tracepoint manager takes the semantic table as the input, and then inserts and activates corresponding tracepoints in the Go binaries under analysis. 
The foundation of \tool{}'s tracepoint manager is DynamoRIO~\cite{dynamorio-www}, which is the state-of-the-art binary analysis engine maintained by Google. DynamoRIO, working on both Linux and Windows, is able instrument x86\_64 and aarch64 instructions, functions, and load modules.

\paragraph{Naive implementation}
As a straightforward scheme, the tracepoint manager utilizes DynamoRIO to overload functions in Go runtime in the semantic table. As the built-in capability, DynamoRIO is able to instrument any function by inserting a pre-function callback at the function entrance to capture the function arguments and a post-function callback at the function exit to capture the function return values and the changed arguments. However, this scheme does not work for Go programs because Go has a different calling convention from that of native C/C++ binaries.   

\paragraph{Issues raised by Go calling convention}
The Go compiler generates a prologue for each Go function to check the stack overflow in the current Goroutine~\footnote{Go functions with the compiler directive ``//go:nosplit" can avoid overflow check but it is not commonly used in Go runtime.}. 
If there is insufficient space on the stack to for the function, Go runtime needs to call {\tt runtime.morestack()} to allocate larger stack space. The Go runtime then moves the function from the old stack to the new space. Such operation triggers an unexpected invocation of the post-function callback earlier than the function exit point. Thus, the tracepoint manager can get a wrong return values of the instrumented Go function. Moreover, if the Goroutine is scheduled back after the {\tt runtime.morestack()} operation, it rolls back the execution to the function entrance, resulting in the pre-function callbacks triggered twice. It is worth noting that this issue occurs not only in DynamoRIO, but also other popular binary instrumentation engines, such as Pin~\cite{Luk-Cohn-etal:2005:PLDI-pin} and Valgrind~\cite{valgrind-www}. Without carefully handling the Go function calling conventions, \tool{} can 
produce incorrect analysis results.

\paragraph{Solution}
We address the issues raised by Go function calling convention by designing a new function instrumentation strategy. For each Goroutine, \tool{} maintains a shadow stack (named sp-stack) that stores all the stack pointers of currently active wrapped functions (i.e., activation records). For each Go function, \tool{} leverages DynamoRIO to instrument its first basic block for function entrance and all the return instructions for function exit. Upon the function entrance, \tool{} checks the current stack point with the top of sp-stack of the Goroutine. If they have the same value, it means that the rollback occurs after the {\tt runtime.morestack()} invocation as the current stack pointer has been already pushed to sp-stack. If it is the case, \tool{} does not trigger the pre-function callback of the current Go function because it is already triggered. Otherwise, it means that the current function is executed with no rollback. \tool{} then pushes the current stack pointer to sp-stack and triggers pre-function callback. 

Upon each return instruction, \tool{} compares the current stack pointer with the top of sp-stack in the same Goroutine. If they are the same, it means the function exit matches the function entrance. \tool{} pops off the top of sp-stack and triggers the post-function callback. Otherwise, it means that the return happens in a callee of the current function. \tool{} just ignores this return and continues the monitoring until captures the return instruction of the current function. This shadow stack approach can always install the function-level tracepoint correctly. Moreover, since sp-stack is per Goroutine, it requires no synchronization, incurring minimum overhead. 

\tool{} uses {\tt goid} to index the sp-stack of each Goroutine, except for {\tt g0}, which is a Goroutine dedicated in each OS thread to execute Go runtime functions within the system stack.  \tool{} maintains sp-stacks of all {\tt g0}s in the OS-supported thread local storage. With this technique, \tool{} is able to manage the tracepoints correctly and efficiently even with massive amounts of Goroutines executed simultaneously. 

\paragraph{Tracepoint installation}
With our function-level instrumentation, \tool{} can install most of the tracepoints and obtain the necessary semantics via function arguments and return values at runtime. Besides function-level tracepoints, \tool{} also supports instruction-level tracepoints, as described in Section~\ref{recovering}. It is worth noting that \tool{} provides extensible APIs to support adding new tracepoints, not only in Go runtime, but also in other native synchronization libraries (e.g., locks, conditional variables, barriers in pthread library) if the program understand analysis is hybridized with Go and other native libraries.


\begin{table*}[t]
\scriptsize
\centering
\resizebox{\textwidth}{!}{%
\begin{tabular}{|c|c|c|c|c|c|c|c|c|c|}
\hline
                       & CockroachDB & etcd & gRPC & Hugo & Istio & Kubernetes & Moby & Knative Serving & Syncthing \\ \hline
Missing Unlock         & 1           & 1    & 2    & 0    & 0     & 0          & 2    & 0               & 0         \\ \hline
WaitGroup              & 2           & 0    & 0    & 0    & 0     & 0          & 1    & 0               & 0         \\ \hline
Channel                & 9           & 1    & 4    & 0    & 2     & 4          & 5    & 0               & 1         \\ \hline
Double Lock            & 4           & 2    & 0    & 2    & 0     & 1          & 1    & 0               & 1         \\ \hline
Mutex Deadlock         & 2           & 0    & 0    & 0    & 0     & 2          & 1    & 0               & 0         \\ \hline
Channel-mutex Deadlock & 0           & 4    & 2    & 0    & 1     & 4          & 1    & 1               & 0         \\ \hline
Cond                   & 0           & 0    & 0    & 0    & 0     & 2          & 2    & 0               & 0         \\ \hline
Total                  & 18          & 8    & 8    & 2    & 3     & 13         & 13   & 1               & 2         \\ \hline
\end{tabular}
}
\caption{The blocking bugs of various categories in real applications collected by GoBench. \tool{} is able to identify all these blocking bugs with no false positives or false negatives.}
\label{tab:evaluation}
\end{table*}

\subsection{Bug Detector}

To detect the blocking bugs, 
\tool{} first leverages the tracepoints to identify all the synchronization primitives and then applies the algorithms described in Section~\ref{methodology}.

\paragraph{Identifying synchronization primitives}
\tool{} monitors both mutex locks and channels. To capture mutex lock creation, acquisition, and release, \tool{} activates the tracepoint at function {\tt runtime.newobject()} in Go runtime, which creates various objects. \tool{} obtains the object type from the argument and the object address from the return value. Such object type information tells whether the object created is a mutex lock and further whether it is a normal lock or a reader-writer lock.
If \tool{} identifies a mutex lock creation, it adds the lock address as the ID to a set (named mutex-set) for later usage. Furthermore, \tool{} activates the tracepoints automatic instructions such as {\tt CMPXCHG} and {\tt XADD} with the {\tt LOCK} prefix to capture lock acquisition and release. 
We cannot directly use the function-level tracepoints for this purpose because some lock and unlock functions are inlined. To filter out atomic instructions that are not related to mutex locks, \tool{} investigates the operands of these instructions. Only the atomic instructions operating on the lock word in the mutex-set are considered lock operations. \tool{} further examines the semantics of the instruction to understand whether it is a lock acquisition or a release. 


\begin{figure*}[t]
\centering
\includegraphics[width=0.8\textwidth]{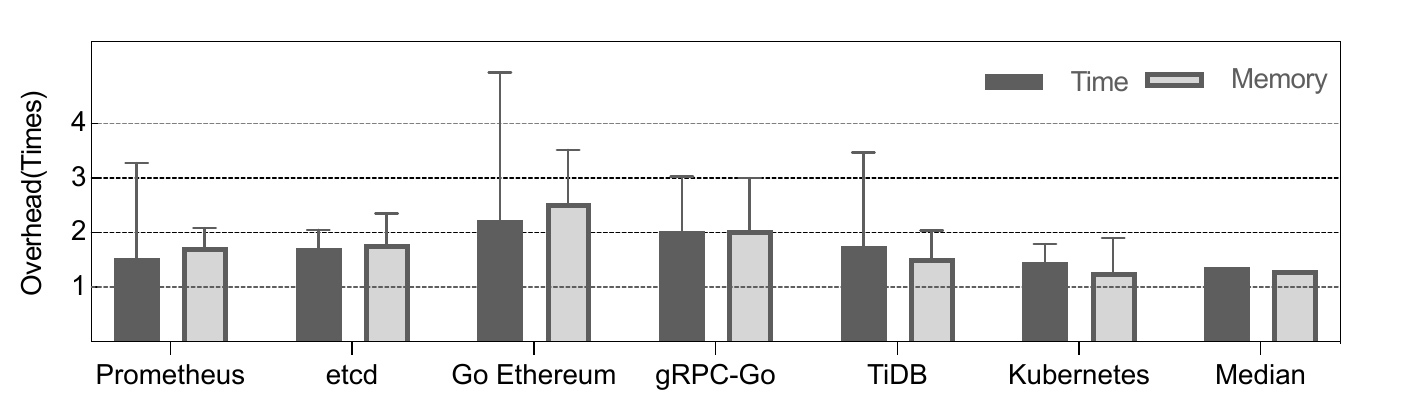}
\centering
\caption{Time and memory overhead of \tool{} on real Go applications. \tool{} incurs moderate overhead in both runtime and memory. Since each real application has hundreds of test inputs, we run all of them and compute the geomean and variance across all of the inputs for an application. We also show the median of runtime and memory overhead across all applications with more than 1,000 executions with different test inputs.}
\label{fig:overhead}
\end{figure*}


\begin{figure}[t]
\begin{center}
\includegraphics[width=0.5\textwidth]{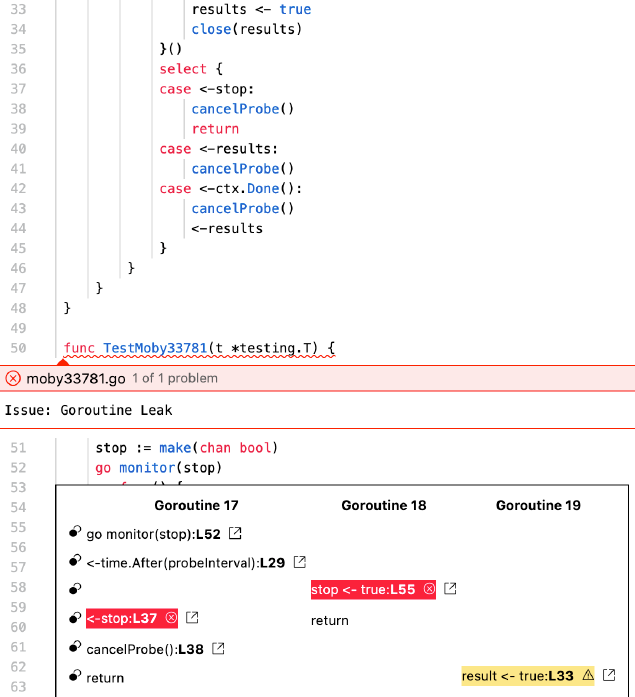}
\end{center}
\caption{The blocking bug involves with both locks and channels in etcd.} 
\label{fig:cs-etcd}
\end{figure}

\tool{} also monitors the channel operations, such as creation, send, receive, and destruction. Go activates the tracepoint at function {\tt runtime.makechan()} to capture the channel creation point. 
\tool{} obtains the return value of this function, which is a pointer to the channel structure {\tt hchan}. \tool{} then decodes {\tt hchan} for the channel ID (using channel address). Moreover, \tool{} activates the tracepoint at function {\tt runtime.closechan()} for closing a channel~\footnote{This function may be not always called to destruct a channel. A channel can be active until the end of the program execution.}. By investigating the function arguments, \tool{} can know the ID of the destructed channel. Furthermore, \tool{} activates function {\tt runtime.chansend()} and {\tt runtime.chanrecv()} to monitor channel {\tt send} and {\tt receive} operations.
Some complexity rises for the semantics of the {\tt select} statement, which schedules multiple active channels (specified in {\tt case} statement) for the current Goroutine. The Go compiler compiles
the {\tt select} statement into the function {\tt runtime.selectgo}. \tool{} can activate the tracepoint at this function to obtain the IDs of all the active channels for selection. \tool{} analyzes all these channels. If one channel can result in blocking, \tool{} reports a blocking bug, which aligns our prediction algorithm described in Section~\ref{methodology}.
Moreover, every channel operation has a boolean argument named {\tt block}, indicating whether the Goroutine can be blocked on this channel operation. \tool{} only investigates channels with its {\tt block} flag set to true and its buffer size set to $0$.


\paragraph{Devising bug analysis}
\tool{} leverages a framework solution for high extensibility. Figure~\ref{fig:overview} illustrates the design principle. \tool{} implements the common features used for analyzing different blocking bugs in the framework. Such common features include activating tracepoints, obtaining lock/channel operations, and associating analysis results with source code. \tool{} builds clients atop the framework to analyze different types of blocking bugs. Each client can customize the use of tracepoints and devise the algorithm as described in Section~\ref{methodology}. If one would like to support a new analysis, he/she only needs to implement a client, which minimizes the coding efforts.

\subsection{Analysis Presentation}
\label{visualization}

As an end-to-end tool, \tool{} needs to present the analysis results in a user-friendly way. We tightly integrate \tool{} into Microsoft Visual Studio Code (vscode)~\cite{vscode-www}, which is one of the most popular code editors for Go. The analysis results, shown as the sequence of synchronization primitives across different Goroutines, are presented in a floating window in the program source code pane. The synchronization primitives involved in the blocking bugs are highlighted and one can easily associate them with source code with a simple mouse click. Figure~\ref{fig:cs-etcd} shows a snapshot of \tool{}'s output, which is described in detail in Section~\ref{etcd}.

\section{Evaluation} \label{evaluation}

We evaluate \tool{} on an x86 machine, which employs a 3.0 GHz 18-core Intel Xeon W-2295 processor, 32 KB private L1 data cache, 1 MB private L2 cache, 25 MB shared L3 cache, and 256 GB Memory. The OS is Ubuntu 20.04 with Linux kernel 5.11.0 and Go 1.17.3. We evaluate \tool{} from two aspects: bug coverage and measurement overhead. 

\paragraph{Bug Coverage}
We evaluate the coverage of \tool{} using \textsc{GoKer} from \textsc{GoBench}~\cite{9370317}, which consists of bug kernels from well-known open-source Go applications. Table~\ref{tab:evaluation} summarizes all the blocking bugs in GoKer. We exclude non-blocking bugs, such as data races, because they are not the target of \tool{}. \tool{} is able to identify all the 68 blocking bugs with no false positives and false negatives. We compare \tool{} with the state-of-the-art blocking bug detector---GCatch~\cite{10.1145/3445814.3446756}. GCatch can detect blocking bugs related to channels and mutex locks only, but not other blocking bugs. Moreover, we find that GCatch shows up a large number of false positives due to its static analysis, which adds additional burdens to users to identify real bugs. 


\begin{table}[t]
\scriptsize
\centering
\renewcommand\arraystretch{1}
\resizebox{.5\textwidth}{!}{%
\begin{tabular}{|l|l|}

\hline

Go applications & Description \\

\hline
Prometheus~\cite{prometheus-www}  & A systems and service monitoring system                                                                                                           \\ \hline
etcd~\cite{etcd-www}        & A distributed key-value store system                                                                                                              \\ \hline
Go Ethereum~\cite{goethereum-www} & Official Go implementation of the Ethereum protocol                                                                                               \\ \hline
gRPC-Go~\cite{grpcgo-www}     & The Go implementation of gRPC                                                                                                                     \\ \hline
TiDB~\cite{TiDB-www}        & An open-source NewSQL database                                                                                                                    \\ \hline
Kubernetes~\cite{kubernetes-www}  & \begin{tabular}[c]{@{}l@{}}An open-source system for automating deployment, \\ scaling, and management of containerized applications\end{tabular} \\ \hline
\end{tabular}
}
\caption{The description of real Go applications used in the evaluation.}
\label{tab:applications}
\end{table}

\paragraph{Overhead Measurement}
We quantify the overhead of \tool{} using real Go applications in Table~\ref{tab:applications}. The overhead is shown in Figure~\ref{fig:overhead}.
The runtime overhead, computed as the execution time of online analysis over the native execution, is typically 1.38$\times$. We also show the overhead variance of each application if it has multiple test inputs. Similarly, we measure the memory overhead, which is defined as the peak memory consumption of the application execution monitored by \tool{} over the peak memory consumption of the native application execution. From the figure, we can see that \tool{} typically incurs 1.32$\times$ memory overhead. The overhead is proportional the number of tracepoints enabled in the execution so the variance for some applications is large. Both runtime and memory overheads are affordable for the production usage.


\section{Case Studies} \label{case_stuties}
We study three Go applications to show how \tool{} can pinpoint blocking of different types and provide intuitive guidance for bug fixing.

\subsection{Channel-Mutex Deadlock in etcd}
\label{etcd}

Figure~\ref{fig:cs-etcd} shows the snapshot of \tool{} that reports a blocking bug in etcd. As described in Section~\ref{visualization}, the analysis results are visualized in vscode. Figure~\ref{fig:cs-etcd} shows the source code pane of vscode, where the source code of etcd is presented. \tool{}'s analysis results are in a floating window. In the window, only problematic Goroutines with their synchronization primitive sequences are shown. In this report, Goroutine 18 and 19 are involved in a blocking bug. The synchronization primitives highlighted in yellow indicate the locations that the blocking occurs. The synchronization primitives highlighted in red are the reason that causes the blocking. One can click the label of any synchronization primitive to jump to the source code line in the vscode source code pane. 

From this figure, we can see that this blocking bug involves both a mutex lock and a channel. The problematic Goroutines 18 and 19 are blocked due to the contention on the same mutex lock at line 38 and line 43, respectively. The channel operations---close at line 29 and receive at line 46---cause the blocking as Goroutine 19 needs to wait for the channel sending or closing events.


\begin{figure}[t]
\begin{center}
\includegraphics[width=0.5\textwidth]{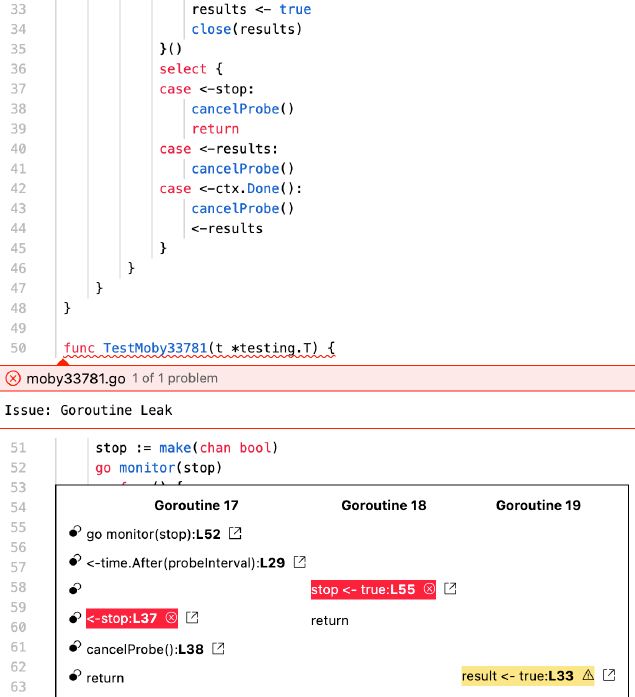}
\end{center}
\caption{A blocking bug due to double RLock in Kubernetes~\cite{kubernetes-www}.}
\label{fig:cs-kb}
\end{figure}

\begin{figure}[t]
\begin{center}
\includegraphics[width=0.49\textwidth]{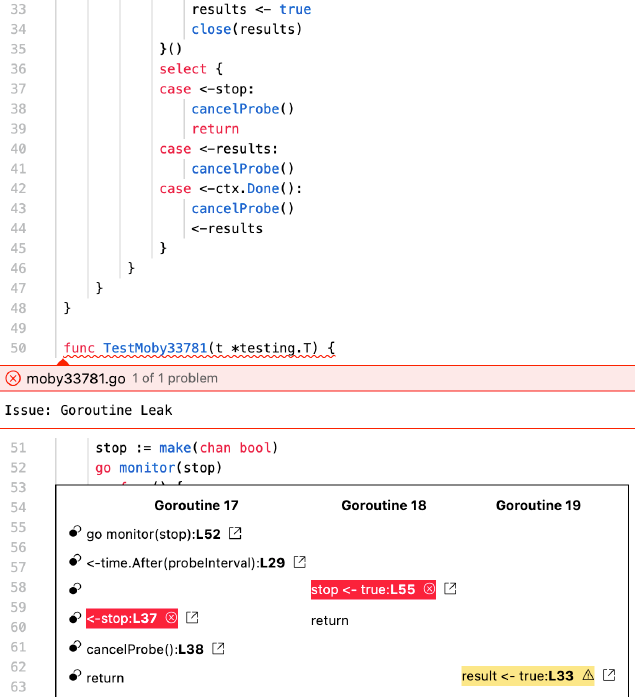}
\end{center}
\caption{A blocking bug due to a blocked channel in Moby~\cite{mobyproject-www}.}
\label{fig:cs-mody}
\end{figure}

\subsection{Double Reader Locks in Kubernetes}

Figure~\ref{fig:cs-kb} shows the snapshot of \tool{} that reports a blocking bug in Kubernetes. From this figure, we can see that the blocking bug is caused by interleaving two reader locks. The problematic Goroutines 17 and 18 are blocked because the writer lock acquisition at line 57 in Goroutine 18 is in between the two reader locks at lines 33 and 42 in Goroutine 17, respectively. The priority of a writer lock is higher than reader locks, resulting in cyclic waiting between the two Goroutines.

\subsection{Blocked Channel in Moby}

Figure~\ref{fig:cs-mody} shows a snapshot of \tool{} reporting a blocking bug in Moby. From the figure we can see that Goroutine 19 is blocked on a channel send operation at line 33. The statements highlighted in red shows that there is no corresponding channel receive operation for the blocked channel send operation. Thus, a blocked channel occurs.

\section{Discussions}
\label{discussion}
In this section, we discuss some limitations of \tool{}. First, \tool{} is a dynamic bug detector, so it cannot detect concurrency bugs in paths that are never executed. This limitation is common to all dynamic analysis tools. To increase the bug coverage, one can explore different inputs (i.e., input fuzzing) or different scheduling strategies (i.e., scheduling fuzzing) to increase the path coverage for analysis. Second, \tool{} requires some manual efforts to identify the interesting tracepoints in Go runtime, which is tedious and error-prone. Fortunately, such efforts are needed only once for each Go release. We have already maintained the tracepoint list for several recent Go versions, which requires no additional efforts from users to use \tool{}. Third, \tool{} requires the availability of the Go application's source code to interpret the root causes of concurrency bugs. It is common for all bug detectors, as the report is usually associated with the source code. However, as a unique advantage, \tool{} does not need the availablity of appliaction source code to run the detection because all the measurement and analysis are done on binary. Lastly, \tool{} is a bug detector, which does not automatically fix the bug like the prior tool~\cite{10.1145/3445814.3446756}. However, \tool{} produces accurate analysis report and its tight integration with vscode provides an opportunity to guide automatic bug fixing.

\section{Conclusions and Future Work} \label{conclusions}

This paper presents \tool{}, a novel bug detector for Go applications. As its unique feature, \tool{} mainly analyzes Go binaries, which is more suitable for deployment in the complex production environment coded with hybrid Go and native languages. \tool{} recovers Go semantics, devises efficient bug detection algorithms, and presents analysis results in user friendly GUI. Our experiments show that \tool{} has high bug coverage, with no false positives or false negatives. Moreover, \tool{} incurs moderate overhead. The case studies show that \tool{} provides intuitive guidance for fixing the Go bugs. \tool{} will be open source upon the paper acceptance.
As our future work, we will extend \tool{} to detect non-blocking bugs in Go applications, such as data races. Moreover, we will explore the techniques to feed the analysis results of \tool{} into a automatic bug fixing tool.

\bibliographystyle{plain}
\bibliography{reference}

\begin{thebibliography}{10}

\bibitem{aresdb-www}
Aresdb.
\newblock \url{https://github.com/uber/aresdb}.

\bibitem{cockroach-www}
Cockroachdb.
\newblock \url{https://www.cockroachlabs.com}.

\bibitem{docker-www}
Docker.
\newblock \url{https://www.docker.com}.

\bibitem{dropbox-www}
Dropbox.
\newblock \url{https://www.dropbox.com}.

\bibitem{dynamorio-www}
Dynamorio.
\newblock \url{https://dynamorio.org}.

\bibitem{dyninst-www}
Dyninst.
\newblock \url{https://www.dyninst.org}.

\bibitem{etcd-www}
etcd.
\newblock \url{https://etcd.io}.

\bibitem{vet}
Go command vet.
\newblock \url{https://pkg.go.dev/cmd/vet}.

\bibitem{go-www}
The go programming language.
\newblock \url{https://go.dev}.

\bibitem{goruntime}
Go runtime.
\newblock \url{https://pkg.go.dev/runtime}.

\bibitem{goleak}
goleak.
\newblock \url{https://github.com/uber-go/goleak}.

\bibitem{goethereum-www}
Gp ethereum.
\newblock \url{https://geth.ethereum.org}.

\bibitem{grpcgo-www}
grpc-go.
\newblock \url{https://github.com/grpc/grpc-go}.

\bibitem{kubernetes-www}
Kubernetes.
\newblock \url{https://kubernetes.io}.

\bibitem{mobyproject-www}
Moby.
\newblock \url{https://mobyproject.org}.

\bibitem{prometheus-www}
Prometheus.
\newblock \url{https://prometheus.io}.

\bibitem{staticcheck}
Staticcheck.
\newblock \url{https://github.com/dominikh/go-tools}.

\bibitem{TiDB-www}
Tidb.
\newblock \url{https://get.pingcap.com/tidb-developer}.

\bibitem{valgrind-www}
Valgrind.
\newblock \url{https://valgrind.org}.

\bibitem{vscode-www}
Visual studio code.
\newblock \url{https://code.visualstudio.com}.

\bibitem{Brumley:2006:TAG:1130235.1130359}
David Brumley et~al.
\newblock Towards automatic generation of vulnerability-based signatures.
\newblock In {\em Proc. of the 2006 IEEE Symp. on Security and Privacy}, SP
  '06, pages 2--16, 2006.

\bibitem{6718069}
Yan Cai and W.K. Chan.
\newblock Magiclock: Scalable detection of potential deadlocks in large-scale
  multithreaded programs.
\newblock {\em IEEE Transactions on Software Engineering}, 40(3):266--281,
  2014.

\bibitem{6114398}
Trevor~E. Carlson, Wim Heirman, and Lieven Eeckhout.
\newblock Sniper: Exploring the level of abstraction for scalable and accurate
  parallel multi-core simulation.
\newblock In {\em SC '11: Proceedings of 2011 International Conference for High
  Performance Computing, Networking, Storage and Analysis}, pages 1--12, 2011.

\bibitem{Chabbi:2012:DTP:2259016.2259033}
Milind Chabbi and John Mellor-Crummey.
\newblock {DeadSpy: A Tool to Pinpoint Program Inefficiencies}.
\newblock In {\em Proceedings of the Tenth International Symposium on Code
  Generation and Optimization}, CGO '12, pages 124--134, New York, NY, USA,
  2012. ACM.

\bibitem{10.5555/3277355.3277436}
Yuxi Chen, Shu Wang, Shan Lu, and Karthikeyan Sankaralingam.
\newblock Applying hardware transactional memory for concurrency-bug failure
  recovery in production runs.
\newblock In {\em Proceedings of the 2018 USENIX Conference on Usenix Annual
  Technical Conference}, USENIX ATC '18, page 837–850, USA, 2018. USENIX
  Association.

\bibitem{gabet_et_al:LIPIcs:2020:13161}
Julia Gabet and Nobuko Yoshida.
\newblock {Static Race Detection and Mutex Safety and Liveness for Go
  Programs}.
\newblock In Robert Hirschfeld and Tobias Pape, editors, {\em 34th European
  Conference on Object-Oriented Programming (ECOOP 2020)}, volume 166 of {\em
  Leibniz International Proceedings in Informatics (LIPIcs)}, pages 4:1--4:30,
  Dagstuhl, Germany, 2020. Schloss Dagstuhl--Leibniz-Zentrum f{\"u}r
  Informatik.

\bibitem{10.5555/2387880.2387902}
Guoliang Jin, Wei Zhang, Dongdong Deng, Ben Liblit, and Shan Lu.
\newblock Automated concurrency-bug fixing.
\newblock In {\em Proceedings of the 10th USENIX Conference on Operating
  Systems Design and Implementation}, OSDI'12, page 221–236, USA, 2012.
  USENIX Association.

\bibitem{10.1145/3093333.3009847}
Julien Lange, Nicholas Ng, Bernardo Toninho, and Nobuko Yoshida.
\newblock Fencing off go: Liveness and safety for channel-based programming.
\newblock {\em SIGPLAN Not.}, 52(1):748–761, jan 2017.

\bibitem{8453195}
Julien Lange, Nicholas Ng, Bernardo Toninho, and Nobuko Yoshida.
\newblock A static verification framework for message passing in go using
  behavioural types.
\newblock In {\em 2018 IEEE/ACM 40th International Conference on Software
  Engineering (ICSE)}, pages 1137--1148, 2018.

\bibitem{10.1145/3341301.3359638}
Guangpu Li, Shan Lu, Madanlal Musuvathi, Suman Nath, and Rohan Padhye.
\newblock Efficient scalable thread-safety-violation detection: Finding
  thousands of concurrency bugs during testing.
\newblock In {\em Proceedings of the 27th ACM Symposium on Operating Systems
  Principles}, SOSP '19, page 162–180, New York, NY, USA, 2019. Association
  for Computing Machinery.

\bibitem{Lin08automaticprotocol}
Zhiqiang Lin, Xuxian Jiang, Dongyan Xu, and Xiangyu Zhang.
\newblock Automatic protocol format reverse engineering through context-aware
  monitored execution.
\newblock In {\em Network and Distributed System Security}, 2008.

\bibitem{10.1145/3037697.3037735}
Haopeng Liu, Guangpu Li, Jeffrey~F. Lukman, Jiaxin Li, Shan Lu, Haryadi~S.
  Gunawi, and Chen Tian.
\newblock Dcatch: Automatically detecting distributed concurrency bugs in cloud
  systems.
\newblock In {\em Proceedings of the Twenty-Second International Conference on
  Architectural Support for Programming Languages and Operating Systems},
  ASPLOS '17, page 677–691, New York, NY, USA, 2017. Association for
  Computing Machinery.

\bibitem{ibs-ispass}
Xu~Liu and John {Mellor-Crummey}.
\newblock Pinpointing data locality bottlenecks with low overheads.
\newblock In {\em Proc. of the 2013 IEEE Intl. Symp. on Performance Analysis of
  Systems and Software}, Austin, TX, USA, April 21-23, 2013.

\bibitem{10.1145/3445814.3446756}
Ziheng Liu, Shuofei Zhu, Boqin Qin, Hao Chen, and Linhai Song.
\newblock Automatically detecting and fixing concurrency bugs in go software
  systems.
\newblock In {\em Proceedings of the 26th ACM International Conference on
  Architectural Support for Programming Languages and Operating Systems},
  ASPLOS 2021, page 616–629, New York, NY, USA, 2021. Association for
  Computing Machinery.

\bibitem{Luk-Cohn-etal:2005:PLDI-pin}
Chi-Keung Luk, Robert Cohn, Robert Muth, Harish Patil, Artur Klauser, Geoff
  Lowney, Steven Wallace, Vijay~Janapa Reddi, and Kim Hazelwood.
\newblock Pin: Building customized program analysis tools with dynamic
  instrumentation.
\newblock In {\em Proc. of the 2005 {ACM SIGPLAN} conference on Programming
  Language Design and Implementation}, pages 190--200, New York, NY, USA, 2005.
  ACM Press.

\bibitem{Narayanasamy:2005:BCR:1069807.1069994}
Satish Narayanasamy, Gilles Pokam, and Brad Calder.
\newblock Bugnet: continuously recording program execution for deterministic
  replay debugging.
\newblock In {\em Proc. of the 32nd Annual Intl. Symposium on Computer
  Architecture}, ISCA '05, pages 284--295, 2005.

\bibitem{10.1145/2892208.2892232}
Nicholas Ng and Nobuko Yoshida.
\newblock Static deadlock detection for concurrent go by global session graph
  synthesis.
\newblock In {\em Proceedings of the 25th International Conference on Compiler
  Construction}, CC 2016, page 174–184, New York, NY, USA, 2016. Association
  for Computing Machinery.

\bibitem{10.1145/3314221.3322484}
Alceste Scalas, Nobuko Yoshida, and Elias Benussi.
\newblock Verifying message-passing programs with dependent behavioural types.
\newblock In {\em Proceedings of the 40th ACM SIGPLAN Conference on Programming
  Language Design and Implementation}, PLDI 2019, page 502–516, New York, NY,
  USA, 2019. Association for Computing Machinery.

\bibitem{loadspy}
Pengfei Su, Shasha Wen, Hailong Yang, Milind Chabbi, and Xu~Liu.
\newblock {Redundant Loads: A Software Inefficiency Indicator}.
\newblock In {\em Proceedings of the 41st International Conference on Software
  Engineering}, ICSE '19, pages 982--993, Piscataway, NJ, USA, 2019. IEEE
  Press.

\bibitem{10.1145/3297858.3304069}
Tengfei Tu, Xiaoyu Liu, Linhai Song, and Yiying Zhang.
\newblock Understanding real-world concurrency bugs in go.
\newblock In {\em Proceedings of the Twenty-Fourth International Conference on
  Architectural Support for Programming Languages and Operating Systems},
  ASPLOS '19, page 865–878, New York, NY, USA, 2019. Association for
  Computing Machinery.

\bibitem{RVN}
S.~Wen, X.~Liu, and M.~Chabbi.
\newblock {Runtime Value Numbering: A Profiling Technique to Pinpoint Redundant
  Computations}.
\newblock In {\em 2015 International Conference on Parallel Architecture and
  Compilation (PACT)}, pages 254--265, Oct 2015.

\bibitem{redspy}
Shasha Wen, Milind Chabbi, and Xu~Liu.
\newblock {REDSPY: Exploring Value Locality in Software}.
\newblock In {\em Proceedings of the Twenty-Second International Conference on
  Architectural Support for Programming Languages and Operating Systems},
  ASPLOS '17, pages 47--61, New York, NY, USA, 2017. ACM.

\bibitem{Xiang:2013:HHO:2451116.2451153}
Xiaoya Xiang, Chen Ding, Hao Luo, and Bin Bao.
\newblock {HOTL}: A higher order theory of locality.
\newblock In {\em Proceedings of the Eighteenth International Conference on
  Architectural Support for Programming Languages and Operating Systems},
  ASPLOS '13, pages 343--356, New York, NY, USA, 2013. ACM.

\bibitem{9370317}
Ting Yuan, Guangwei Li, Jie Lu, Chen Liu, Lian Li, and Jingling Xue.
\newblock Gobench: A benchmark suite of real-world go concurrency bugs.
\newblock In {\em 2021 IEEE/ACM International Symposium on Code Generation and
  Optimization (CGO)}, pages 187--199, 2021.

\bibitem{10.1145/1950365.1950395}
Wei Zhang, Junghee Lim, Ramya Olichandran, Joel Scherpelz, Guoliang Jin, Shan
  Lu, and Thomas Reps.
\newblock Conseq: Detecting concurrency bugs through sequential errors.
\newblock In {\em Proceedings of the Sixteenth International Conference on
  Architectural Support for Programming Languages and Operating Systems},
  ASPLOS XVI, page 251–264, New York, NY, USA, 2011. Association for
  Computing Machinery.

\end{thebibliography}

\end{document}